\documentclass[11pt,a4paper]{article}
\usepackage{cite}
\usepackage{amsmath, amsthm, amssymb,slashed}
\usepackage{yfonts}

\usepackage{ifpdf}
\ifpdf
  \usepackage[pdftex]{graphicx}
  \usepackage{epstopdf}
\else
  \usepackage[dvips]{graphicx}
\fi

\usepackage{fancyhdr}
\usepackage{mathbbol}
\usepackage{url}
\usepackage{color}
\usepackage{bbm}
\usepackage{rsfso}
\usepackage{dsfont}
\usepackage{bm}
\usepackage{amsfonts}
\usepackage{mathrsfs}
\usepackage{slashed}
\usepackage{wasysym}
\usepackage{yfonts}
\usepackage{ytableau}
\usepackage{tikz}
\usetikzlibrary{arrows,decorations.pathmorphing,hobby}
\usetikzlibrary{calc,arrows.meta}
\usepackage{cancel}
\usepackage[normalem]{ulem}
\usepackage[hidelinks]{hyperref}
\usepackage{subfig}
\usepackage{tabu}

\def\bl{\begin{equation}\begin{aligned}}
\def\el{\end{aligned}\end{equation}}
\def\beal{\begin{align}}
\def\eal{\end{align}}
\def\be{\begin{equation}}
\def\ee{\end{equation}}
\def\bpm{\begin{pmatrix}}
\def\epm{\end{pmatrix}}
\def\bsm{\begin{bmatrix}}
\def\esm{\end{bmatrix}}
\def\bvm{\begin{vmatrix}}
\def\evm{\end{vmatrix}}
\def\bVM{\begin{Vmatrix}}
\def\eVM{\end{Vmatrix}}
\def\bea{\begin{eqnarray}}
\def\eea{\end{eqnarray}}

\newcommand{\braket}[1]{\langle #1 \rangle}

\def\1{{\bf 1}}
\def\2{{\bf 2}}
\def\3{{\bf 3}}
\def\4{{\bf 4}}
 
 \def\SU{{\mathrm{SU}}}
 
\def\tilde{\widetilde}

\usepackage{indentfirst}

\usepackage{mathtools}

\title{A Simple Model of Pentaquarks}
\date{\today}
\author{D. Germani, F. Niliani and  A.D. Polosa\\   {\it $^*$Sapienza University of Rome and INFN, Piazzale Aldo Moro 2, I-00185, Italy}   }
\begin{document}
\maketitle

\begin{abstract}

We describe pentaquarks as `baryo-charmonia'  with  a color octet $c\bar c$   core bonded to a color octet three-quark system.  Fermi statistics of the light quark cloud allows to describe two pentaquark triplets: a lower one, well supported by experiment, and a higher one with strangeness. 
For the time being, the lowest line of the strange triplet has been experimentally identified in a $3\sigma$ peak. Data also suggest  two different production mechanisms for pentaquarks. We show how this can be described  in the proposed scheme. 
\end{abstract}
%
\section{Introduction}
Let the pentaquarks be formed by three light quarks in color octet $(qqq)_{\bm 8}$ orbiting in the mean color field of a charm-anticharm  heavy  pair  $(c\bar c)_{\bm 8}$, a sort of `baryo-charmonium'.  
There are other articles that adopt a similar approach: in \cite{bopenta}, the Born-Oppenheimer approximation is used to study the internal dynamics of the constituents; in \cite{xc}, the color component $\bm 1 - \bm 1$ is also considered~\footnote{We do not consider the case of color singlets, as done in \cite{xc}, because we assume that pentaquarks can result only from quark color forces. Namely, we assume here that the lightest narrow states are $\bm8-\bm 8$ and the mixing with $\bm 1-\bm 1$, if any, has to be negligible.} and the mass spectrum of pentaquarks is calculated in a mass splitting model. Another work on pentaquarks as compact particles is \cite{mit}, where the MIT bag model is adopted. In Sec.~\ref{sec:summary}, we will highlight some of the substantial differences between our work and those previously mentioned.

Besides the compact model, there is an alternative literature in which pentaquarks are described as meson-baryon molecules \cite{molecola,molecola1,molecola2,molecola3,molecola4}.

The Fermi statistics of the light quarks leads to a determination of the spectrum of the best ascertained $J/\psi p$ pentaquarks, $P_c(4312)$, $P_c(4440)$ and $ P_c(4457)$~\cite{gp}, as well as to the prediction of two extra lines  in the strange sector, in addition to the observed one.
The lighter state in the strange pentaquark system is a $3\sigma$ peak, dubbed $P_{cs}(4459)$ by LHCb~\cite{3si}.  The two heavier ones we predict, $P_{cs}^\prime,P_{cs}^{\prime\prime}$,  see Fig.~\ref{pred}, have roughly a similar level of significance and are found in a region where present data show fluctuations over the background. 
For both triplets we predict the same  ordering of spins, namely  $J=1/2,3/2,1/2$, for increasing masses. 

The three $P_c$ pentaquarks are observed in  $\Lambda_b^0\to (J/\psi p) K^-$ decays and the $P_{cs}(4459)$
peak is found in $\Xi_b^-\to (J/\psi \Lambda) K^-$.  Data suggest at least two different production mechanisms for pentaquarks, idependently on their strangeness content. 

In addition to the associated production with $K$ in heavy baryon decays, the $\tilde P_c(4337)$ has been reported by~\cite{ng1} in the decay $B_s^0\to (J/\psi p)\bar p$.  A strange partner of $\tilde P_c$, the $\tilde P_{cs}(4338)$,  is found  in~\cite{ng2} in associated production with the anti-proton ($B^-\to (J/\psi\Lambda)\bar p$).
We use the tilde to distinguish the  pentaquarks produced in association with the anti-proton from those produced in association with the $K$.  We will show how we can describe this  pattern comprising $P$ and $\tilde P$ pentaquarks and place the observed states 
$\tilde P_{c}(4337)$, $\tilde P_{cs}(4338)$ in multiplets with their expected partners, which we  name 
$\tilde P_c^\prime, \tilde P_c^{\prime\prime}$ and $\tilde P_{cs}^\prime, \tilde P_{cs}^{\prime\prime}$.
\begin{table}[t]
    \centering
    \begin{tabular}{|c|c|c||l|c|}
    \hline
         State & Mass [MeV] & Width [MeV]  & Observed Process & Year \\
    \hline\hline
         $P_c(4312)$ & $4311.9\pm0.7^{+6.8}_{-0.6}$ & $9.8\pm2.7^{+3.7}_{-4.5}$ & $\Lambda^0_b\to(J/\psi\,p)\,K^-$ & 2019\\
         $\tilde P_c(4337)$ & $4337^{+7\,\, +2}_{-4\,\, -2}$ & $29^{+26\,\,+14}_{-12\,\,-14}$ & $B^0_s\to(J/\psi p)\,\,\overline{p}$ & 2022\\
         $P_c(4440)$ & $4440.3\pm1.3^{+4.1}_{-4.7}$ & $20.6\pm4.9^{+8.7}_{-10.1}$ & $\Lambda^0_b\to(J/\psi\,p)\,K^-$ & 2019\\
         $P_c(4457)$ & $4457.3\pm0.6^{+4.1}_{-1.7}$ & $6.4\pm2.0^{+5.7}_{-1.9}$ & $\Lambda^0_b\to(J/\psi\,p)\,K^-$ & 2019\\
         $\tilde P_{cs}(4338)^{\frac{1}{2}^-}$ & $4338.2\pm0.7\pm0.4$ & $7.0\pm1.2\pm1.3$ & $B^-\to (J/\psi\,\Lambda)\,\overline{p}$ & 2022\\
         $P_{cs}(4459)$ & $4458.9\pm2.9^{+4.7}_{-1.1}$ & $17.3\pm6.5^{8.0}_{-5.7}$ & $\Xi_b^-\to(J/\psi\,\Lambda)\,K^-$ & 2021 \\
    \hline
    \end{tabular}
    \caption{Pentaquarks discovered by the LHCb collaboration \cite{gp,3si,ng1,ng2}. The first 4 states have light quarks content $uud$, the last two have $uds$. For $P_{cs}^0(4338)$, the experimentally preferred $J^P$ is indicated next to the name. }
    \label{tab:pentaquarks}
\end{table}
\section{Fermi statistics in baryo-charmonia}
The light quarks carry color, in the adjoint representation, flavor, spin and orbital quantum numbers and are identical particles obeying Fermi statistics. Let $a,b,c$ be flavor indices and $\alpha,\beta,\gamma$  color indices. Requiring both color and flavor to be in the adjoint representation we can form the tensor
\be
A^{ijk}_{\alpha\beta\gamma}=\psi^{[a}_\alpha\psi_\beta^{b]}\psi_\gamma^c-\psi^{[a}_\beta\psi_\gamma^{b]}\psi_\alpha^c= 
\psi^{a}_{(\alpha}\psi_{\beta)}^{b}\psi_\gamma^c-\psi^{a}_{(\beta}\psi_{\gamma)}^{b}\psi_\alpha^c\,,
\label{antisimmex}
\ee
which is  anti-symmetric  under the exchange of any two  quarks provided that 
\be
\psi^{a}_\alpha\psi^{b}_\beta=-\psi^{b}_\beta\psi^{a}_\alpha\,.
\label{antix}
\ee
Parentheses in~\eqref{antisimmex} indicate symmetrization (round brackets) or anti-symmetrization (square brackets) of a certain pair of indices. The tensor $A$ has the symmetries given by the following Young Tableaux  in the color $(\bm 8_C)$ and flavor $(\bm 8_F)$ spaces respectively 
\be
{\begin{ytableau}
       \alpha & \beta  \\
  \gamma & \none
\end{ytableau}}\qquad
{\begin{ytableau}
       a & c  \\
  b & \none
\end{ytableau}}\,.
\notag
\ee
If we first symmetrize $\psi_\alpha^a\psi_\beta^b\psi_\gamma^c$ with respect to $\alpha\beta$ and then  antisymmetrize the result with respect to $\alpha\gamma$ (YT on the left) we obtain the middle term Eq.~\eqref{antisimmex}. 
If we first symmetrize $\psi_\alpha^a\psi_\beta^b\psi_\gamma^c$ with respect to $ac$  and then  antisymmetrize the result with respect to $ab$ (YT on the right) we obtain  the right hand side of~\eqref{antisimmex}. 

Similarly the tensor $S$ can be formed 
\be
S^{abc}_{\alpha\beta\gamma}=\eta_{\alpha}^{[a}\eta_\beta^{ b]}\eta_{\gamma}^c
-\eta_{\beta}^{[a }\eta_\gamma^{ b]}\eta_{\alpha}^c
= \eta^a_{\left[\alpha\right.}\eta^b_{\left.\beta\right]}\eta^c_\gamma-\eta^a_{\left[\beta\right.}\eta^b_{\left. \gamma\right]}\eta^c_\alpha   \label{simmex}
\ee
with 
\be
\overline
{\begin{ytableau}
       \alpha & \beta  \\
  \gamma & \none
\end{ytableau}}\qquad
{\begin{ytableau}
       a & c  \\
  b & \none
\end{ytableau}}\,,
\notag
\ee
the bar on the table  indicating that we first anti-symmetrize with respect to  $\alpha\beta$ and then symmetrize with respect to  $\alpha\gamma$.
This is symmetric under the exchange of any two quarks, provided that
\be
\eta^{a}_\alpha\eta^{b}_\beta=+\eta^{b}_\beta\eta^{a}_\alpha
\label{twist}\,.
\ee

Let us derive the standard baryon octet from $S^{abc}_{\alpha\beta\gamma}$ . Assume for the moment that tensor $S$ in~\eqref{simmex} represents the flavor-spin configuration of three quarks in the $(\bm 8_F)$ baryon octet, so that $\alpha,\beta,\gamma$ are spin indices rather than color indices. Assign the additional color indices $i,j,k$ to quarks: $i$ in association with  $(a,\alpha)$, $j$ with $(b,\beta)$ and $k$ with $(c\gamma)$. Anti-symmetrize indices $i,j,k,$ (so that we can eventually contract with $\epsilon_{ijk}$ in order to form a color-singlet). From~\eqref{twist} and anti-symmetrizing $ij$
\be
\eta^{ai}_\alpha\eta^{bj}_\beta=\eta^{bi}_\beta \eta^{aj}_\alpha=-\eta^{bj}_\beta\eta^{ai}_\alpha
\ee
so that we can use the symbol $\eta\to \psi$, as in~\eqref{antix}. 
Symmetry in spin-flavor space combined with anti-symmetry in color space gives Fermi statistics and the standard baryon octet. Indeed contracting $S$ with the Levi-Civita $\epsilon_{ijk}$ to form the singlet gives
\be
{\cal S}^{abc}_{\alpha\beta\gamma}=\epsilon_{ijk} (\psi^{ai}_{\left[\alpha\right.}\psi^{bj}_{\left. \beta\right]}\psi^{ck}_\gamma+\psi^{ai}_{\left[\gamma\right.}\psi^{bj}_{\left.\beta\right]}\psi^{ck}_\alpha)= B^{acb}_{\alpha\gamma\beta}
\label{stba}
\ee
where the baryon octet is indeed usually reported in the form\footnote{The baryon  flavor octet is usually reported as
\be
B^{abc}_{\alpha\beta\gamma}=\epsilon_{ijk} (M^{abc})^{ijk}_{\alpha \beta \gamma}\,,
\ee
where $abc$ are flavor indices, $ijk$ are color indices and $\alpha\beta\gamma$ are spin indices. The symbol $M^{abc}$ corresponds to symmetrizing/antisymmetrizing  the product $\psi^{ai}_\alpha\psi^{bj}_\beta \psi^{ck}_\gamma$, in the flavor space $abc$, as prescribed by the tableaux   
{\tiny $
{\begin{ytableau}
       a & b  \\
  c & \none
\end{ytableau}}
$.} What is found in~\eqref{stba} corresponds to ${\cal S}^{abc}_{\alpha\beta\gamma}=B^{acb}_{\alpha\gamma\beta}$, i.e. to a simultaneous renaming of indices $b\leftrightarrow c$ and $\beta \leftrightarrow \gamma$ in $B$. 
}
$B^{abc}_{\alpha\beta\gamma}$ (for example the proton is $p=B^{121}_{\alpha\beta\gamma}$ the neutron is $n=B^{122}_{\alpha\beta\gamma}$ and so on). 
The property $B^{abc}_{\alpha\beta\gamma}=-B^{bac}_{\alpha\beta\gamma}$ allows to construct the dual $B^c_d=1/2\, \epsilon_{abd}B^{abc}$ which can be represented in the well known  baryon matrix form. The anti-symmetric pairings  $[\alpha,\beta]$ and $[\gamma,\beta]$ correspond to the $j=0$ representation of SU(2), so that each baryon component is in the $j=1/2$ representation of SU(2): the baryon octet has spin $1/2$. 

Coming  back to  pentaquarks, differently from baryons,   the $qqq$ system can either be in a color-flavor state $A$, Eq.~\eqref{antisimmex}, or in $S$, Eq.~\eqref{simmex}.  In the first case the spin-orbital state must be symmetric whereas in the second case it must be 
anti-symmetric, to enforce Fermi statistics. \\
In our brief discussion on baryons we took $S$ and antisymmetrized the color labels. In pentaquarks  light quarks are  in a flavor octet, like in  baryons, but can have both spins $1/2$ and $3/2$, differently from octet baryons, since they are not in an antisymmetric color configuration like~\eqref{stba}.

\begin{figure}
    \centering
    \includegraphics[width=0.5\textwidth]{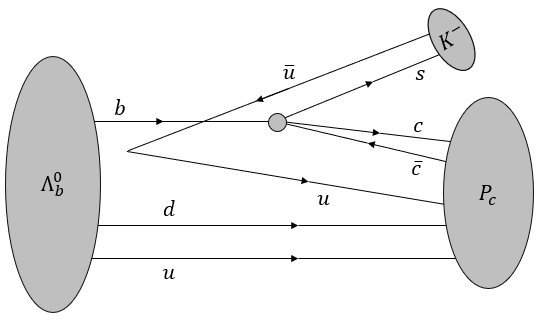}
    \caption{Possible diagram for the decay $\Lambda^0_b\to P_c\,K^-$.}
    \label{fig:decay}
\end{figure}

Consider pentaquarks produced in the decay of a baryon in association with a $K^-$ meson. In the $\Lambda^0_b$ baryon, the $ud$ diquark is in the antisymmetric spin zero state, symmetric in color-flavor space. This is called the ``good" diquark, as opposite to the spin one bad diquark in the $\Sigma_Q$. In the simplest decay process, the initial light quarks propagate to the final state as in Fig.~\ref{fig:decay}. Assuming that the color-flavor symmetry of the $ud$ pair is maintained in the formation of the final state, we choose  $S$ in Eq.~\eqref{simmex}, for the description of the light quarks in the pentaquark. $\Xi_Q$ baryons carry good diquarks as well.
The different production mechanism of the $\tilde P$ pentaquarks leads us to distinguish them from the $P$ pentaquarks by using  
$A$, as in Eq.~\eqref{antisimmex}, in place of $S$.  This will be discussed in Section~\ref{ptilde}.  

Each quark pair, say quark 1 and quark 2,  can either be in a anti-symmetric $S=0$ state or in a symmetric $S=1$ state. The orbital wave-function of the  quark pair (we will call $\Psi$ the spin wave-function, and $\Phi$ the orbital one) will be, accordingly, symmetric or anti-symmetric 
\be
\Phi(\bm r_1,\bm r_2)=\frac{1}{\sqrt{2}}\left( \phi_1(\bm r_1)\phi_2(\bm r_2)\pm \phi_1(\bm r_2)\phi_2(\bm r_1)\right)\,,
\ee
where $\bm r_{1,2}$ are centered in the $c\bar c$ core.

\section{Exchange interaction}
Let $U$ be the color interaction potential between, say,  quark $1$ and quark $2$. The expectation value 
\be
\braket{U}_{\Phi} =\int \Phi^*(\bm r_1,\bm r_2)\, U(\bm r_1-\bm r_2)\, \Phi(\bm r_1,\bm r_2)\, d^3 r_1\, d^3 r_2
\ee
can be written as $C\pm J$, with the $\pm$ signs corresponding to $\Phi$ being symmetric/anti-symmetric. This in turn can be written as $\braket{U}_{\Phi}=C+\braket{V}_\Psi$ provided that $\braket{V}_\Psi=\pm J$ depending on $\Psi$ being anti-symmetric ($S=0$) or symmetric ($S=1$) respectively. The potential  $V$ is given by\cite{Ashcroft}
\be
V=-\sum_{\rm pairs} J_{ab}\left(\frac{1}{2}+2 \bm S_a\cdot \bm S_b\right)
\label{exchange}
\ee
where
\be
J_{ab}=\int (\phi_a(\bm r_a)\phi_b(\bm r_b) )^*\,  U(\bm r_a-\bm r_b)\, (\phi_a(\bm r_b) \phi_b(\bm r_a))\, d^3r_a\, d^3r_b\,.
\ee

Using the basis of states $|-++\rangle,|+-+\rangle,|++-\rangle$ (in Appendix~\ref{app:6jwigner}, we provide a demonstration of the following equations using 6j-Wigner symbols) one obtains that $\langle V\rangle_\Psi$ splits the two spin $1/2$ states, obtained by the combination of three spins $1/2$, by
\be
\Delta E_{1/2}=\pm\sqrt{J_{12}^2+J_{13}^2+J_{23}^2-J_{12}J_{13}-J_{12}J_{23}-J_{13}J_{23}}\,.
\label{eq:shift_1/2}
\ee
The spin $3/2$ shift is readly obtained by $|+++\rangle$ (or $|---\rangle$) to be
\be
\Delta E_{3/2}=-J_{12}-J_{13}-J_{23}\,.
\label{eq:shift_3/2}
\ee
In the case of baryons, where a full color anti-symmetry holds, we have $J_{12}=J_{13}=J_{23}=J$ so that $\Delta E_{1/2}=0$ and, as commented above, there is no $3/2$ spin.

Orbital wave functions are not known. As for the color potential $U$, we might use the one-gluon exchange interaction concluding that if the $i,j$ quark pair were in a color-symmetric configuration we would get a positive, repulsive coupling $J_S$, which, in modulus, is half  the negative coupling $J_A$ of the color anti-symmetric configuration~\footnote{The quadratic Casimir in the repulsive, symmetric, $\bm 6$ representation is $C(\bm 6)=10/3$ so that $2(C(\bm 6)-2C(\bm 3))=-(C(\bar{\bm 3})-2C(\bm 3)))$.}
\be
J_S=-\frac{1}{2}J_A\label{fo2}\,.
\ee  

Let us consider the case $qqq=uud$ in the color-flavor configuration $S$. Then from~\eqref{simmex}
\be
S^{121}_{\alpha\beta\gamma}= u_{\left[\alpha\right.}d_{\left.\beta\right]}u_\gamma+u_{\left[\gamma\right.}d_{\left. \beta\right]}u_\alpha  
\ee
since the $uu$ cannot be anti-symmetric in flavor space. The pentaquark contains the light  quarks $u_\alpha,d_\beta$ and $u_\gamma$ with the $u$ quarks, $u_\alpha$ and $u_\gamma$,  in a color symmetric (repulsive) representation and $ud$   in  color attractive, anti-symmetric pairings~\footnote{
The $c\bar c$ core has color $c^A\bar c_D\epsilon^{DBC}=M^{ABC}$ and the color neutral pentaquark is obtained by
\be
P^{uud}\propto M^{ABC}\,\epsilon_{(\alpha\beta A}\, \epsilon_{\gamma)BC}\, S_{\alpha\beta\gamma}^{121}
\ee
symmetrizing the $\alpha\gamma$ pair as in $S$}. 
Therefore from~\eqref{fo2} we require
\be
J^{uu}_S=-\frac{1}{2}J_A^{ud}
\ee
and we can write
\be
J_{12}=J^{uu}_S>0 \qquad J_{13}=J_{23}=J_A^{ud}<0\,.
\label{conds}
\ee

The pentaquarks discovered in the $J/\psi p$ channel, from $\Lambda_b$ decays, are found at mass values \cite{gp}
\bea
M_{P_c}(4312)&=& 4311.9^{+7}_{-0.9}~\text{MeV}\,,\notag\\
M_{P_c}(4440)&=& 4440^{+4}_{-5}~\text{MeV}\,,\\
M_{P_c}(4457)&=& 4457.3^{+7}_{-1.8}~\text{MeV}\,.\notag
\eea
The spins are not known so far. Assume that the ordering in mass corresponds to the lower one being spin $1/2$ and the higher two being $3/2$ and $1/2$ respectively. Then we have to solve the simultaneous equations
\bea
M_{P_c}(4457)-M_{P_c}(4312)&=&2|J_S^{uu}-J_A^{ud}|\,,\\
M_{P_c}(4440)-\frac{1}{2}(M_{P_c}(4312)+M_{P_c}(4457))&=& -J_S^{uu}-2J_A^{ud} 
\eea
The first equation corresponds to the $\Delta E_{1/2}$ splitting in~\eqref{eq:shift_1/2} with the conditions~\eqref{conds} and the second to the $\Delta E_{3/2}$ shift in~\eqref{eq:shift_3/2}.
This system of equations has two sets of solutions, but only one is compatible with the condition of having one positive and two negative couplings. These are found to be
\bea
J_S^{uu} &=&29.9^{+2.5}_{-2.8}~\text{MeV}\,,\notag\\
J_A^{ud}&=&-42.8^{+2.4}_{-1.6}~\text{MeV}\,,
\label{jud}
\eea
which gives 
\be
\frac{J_S^{uu}}{ J_A^{ud}}=-0.7\pm 0.1\,,
\label{rapp}
\ee
not far from the  $-1/2$ factor in~\eqref{fo2} we aimed to.\\
To conclude, let's define $M_0$ as the degenerate mass of the triplet, i.e., the mass that the three particles analyzed in this section would have if we turned off the exchange interactions. The value of $M_0$ is given by the average mass of the particles with spin $1/2$.
\be
    M_0=\frac{1}{2}(M_{P_c}(4312)+M_{P_c}(4457))=4384^{+4}_{-1}\,\text{MeV}\,.
\ee
If we consider the spin ordering  $1/2,1/2,3/2$ for increasing mass values, we would get $S/A\simeq -0.32^{+0.05}_{-0.07}$. This gives a  preference to~\eqref{rapp} and to the `inverted' spin ordering $1/2,3/2,1/2$ which we will apply also to the strange pentaquarks in the next section.

We will take this $S/A\simeq -0.7$ ratio as a benchmark in pentaquarks (also in the case of strange pentaquarks) and assume  that $J^{ud}=J^{uu}=J^{dd}\equiv J^{qq}$ in both  $A$ or $S$ symmetries.
\section{Strange pentaquarks}
In addition to the three pentaquark lines described above, a strange $J/\psi\Lambda$ pentaquark has been discovered, with a mass value of \cite{3si}
\begin{equation}
    M_{P_{cs}}(4459)=4458.8^{+6}_{-3.1}~\text{MeV}\,.
\end{equation}
Its mass difference with $M_{P_c}(4312)$ is approximately equal to the baryon mass difference $ M(\Lambda)-M(p)$. This suggests to assume that $P_{cs}$ is also a spin $1/2$ state, and like the  $P_c^+$, it is the first of a higher strange triplet. In the following we will determine the  triplet with strangeness extending the analysis done above.

Along the same lines we consider the $\Lambda$-like color-flavor symmetric  combination 
\begin{equation}
    S^{123}_{\alpha\beta\gamma}=u_{[\alpha}d_{\beta]}s_\gamma+u_{[\gamma}d_{\beta]}s_{\alpha}\,.
    \label{x25}
    \end{equation}    
From this we can infer that $u$ and $s$ are in a symmetric, repulsive, pairing $J^{us}_S$ and $u,d$ are in an anti-symmetric pairing $J^{ud}_A$, which will be taken from~\eqref{jud}. As for $ds$, we can symmetrize and anti-symmetrize  color indices so that $J^{ds}= (J_S^{us}+J_A^{us})/2$ where we assume $J_{S,A}^{us}= J_{S,A}^{ds}=J_{S,A}^{qs}$. We will derive $J_A^{us}$ from  $J_S^{us}$ using the same $S/A=1/k$ ratio given  in~\eqref{rapp} so that 
\be
J^{ds}=\frac{1+k}{2}J^{qs}_S\,,
\ee
where $k\simeq - 1.43$, i.e. the inverse of the number in~\eqref{rapp}. 

\begin{figure}[htb!]
  \centering
  \includegraphics[width=0.7\textwidth]{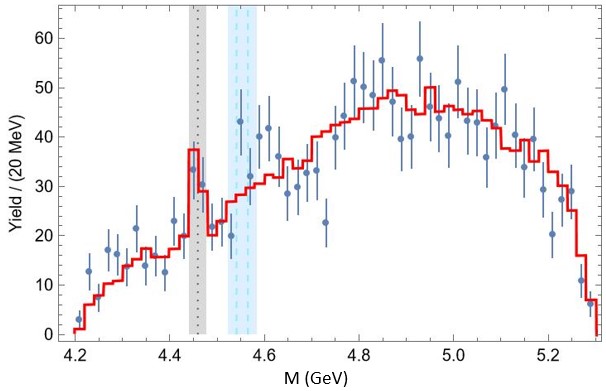}
  \caption{ The dotted line corresponds to the $J/\psi\Lambda$ resonance $P_{cs}(4459)$ reported by LHCb.  We take the fitting (red) curve to the $P_{cs}$ peak from LHCb as well~\cite{3si}. The dashed lines, with uncertainty bands, correspond to our predictions for $P_{cs}^\prime$ and $P_{cs}^{\prime\prime}$. The state $P_{cs}^{\prime\prime}$ is about  20~MeV below the $\bar{D}^{*}\Xi^{\prime}_c$ neutral or charged threshold (which would give a radius $r\sim1/\sqrt{2\mu B}\sim 0.96$~fm, compatible with a compact state). For the two predicted lines, there are no other nearby molecular thresholds with the right quantum numbers. }
  \label{pred}
\end{figure}

We know from data on baryons that 
the ratio of chromomagnetic couplings in the constituent quark model is $\kappa^{qs}_A/\kappa^{qq}_A\sim 0.6$~\cite{Ali:2019roi}. Applying the same scaling law to $J_A$'s, and consequently to $J_S$'s, we have 
$J_S^{qs}=17.9\pm 2 $~MeV which leads to $J^{ds}=-3.9\pm 2$~MeV.
The splitting formulae then give
\begin{gather}
    \Delta E_{1/2}=53.3\pm 2.3\,\text{MeV}\,,\\
    \Delta E_{3/2}=28.7\pm 3.5\,\text{MeV}\,.
\end{gather}
The degenerate mass of this new triplet is
\be
     M_0^s=M_{P_{cs}}+\Delta E_{1/2}=4512\pm 6\,\text{MeV}\,,
\ee
with $\Delta_s=M_0^s-M_0=(127\pm 6)\,$ MeV to be compared to the analogous $\Delta_s\simeq 177$~MeV known from the baryon octet.\\
The mass spectrum is 
\bea
    M_{P'_{cs}(3/2)}&=&M_0^s+\Delta E_{3/2}=4541\pm6\,\text{MeV}\,,\notag\\
    M_{P''_{cs}(1/2)}&=&M_0^s+\Delta E_{1/2}=4565\pm 6\,\text{MeV}\,.
\eea
Therefore the strange pentaquark spectrum is superimposed  on available data in Fig.~\ref{pred}. The predicted states end up in a region of $\sim 2\sigma$ fluctuations over the background, so that it is difficult to make any definite conclusion different from an hint to look better at this region. 

\section{Pentaquarks produced in association with anti-protons\label{ptilde}}
There are two more pentaquarks reported by  experiment as of today. These are found at the mass values \cite{ng1,ng2}
\begin{equation}
    \begin{split}
        &M_{\tilde P_c}(4337)=4337^{+7}_{-4}~\text{MeV}\,,\\
        &M_{\tilde P_{cs}}(4338)=4338.2\pm 0.8~\text{MeV}\,.        
    \end{split}
\end{equation}
We use the $\tilde P$ notation  to underscore the fact that these two resonances have a different  production mechanism  with respect to the pentaquarks discussed above. For example, $\tilde P_{cs}$ is found in $B^-\to (J/\psi \Lambda)\bar p$, differently from its $P_{cs}$ partner observed in $\Lambda_b\to (J/\psi \Lambda)K $. 
The  $\tilde P_{cs}(4338)$ pentaquark, differently from all other states, has also an experimental  spin assignment, namely $J^P=\frac{1}{2}^-$. The $\tilde P_c(4437)$ is found in $B_s^0\to (J/\psi p)\bar p$. We attempt a description of these two pentaquarks as belonging to a different spectroscopic series, which in our picture is characterized by the antisymmetric $A$ color flavor configuration~\eqref{antisimmex}.

Consider the  $\tilde P_{cs}(4338)$. For this we have an experimental determination of $J=\frac{1}{2}$. We consider that the color-flavor state is described as in~\eqref{antisimmex}
\begin{equation}
    A^{uds}_{\alpha\beta\gamma}=u_{(\alpha}d_{\beta)}s_\gamma-u_{(\gamma}d_{\beta)}s_\alpha\,,
\end{equation}
with $A^{uds}_{\alpha\beta\gamma}=-A^{uds}_{\gamma\beta\alpha}$. The spin-orbit in this case works in the opposite way with respect to what we did above: a quark pair with $S=0$  has to have an anti-symmetric orbital, i.e. $\braket{V}_\Phi=-J$. So the overall sign of the $V$ in~\eqref{exchange} changes to positive. Following the same steps as before we have: $J^{ud}_S,J^{us}_A, J^{ds}$ and we will use the same $k=A/S$ factor. Adopting the same couplings we already computed it is found
\begin{equation}
    \begin{split}
        &\Delta E_{3/2}=(J^{ud}_S+J^{us}_A+J^{ds})=\pm4~\text{MeV}\,, \\
        &\Delta E_{1/2}=48.5\pm 2.1~\text{MeV}\,.
    \end{split}
\end{equation}
Given that $\widetilde{M}_0^s=M_{\Tilde{P}_{cs}}+\Delta E_{1/2}=4386\pm2.2\,\text{MeV}$, the two additional states in the spectrum would be
\bea
    M_{\Tilde{P}_{cs}'(3/2)}&=&\widetilde{M}_0^s+\Delta E_{3/2}=4387\pm4\,\text{MeV}\,,\notag\\
    M_{\Tilde{P}_{cs}''(1/2)}&=&\widetilde{M}_0^s+\Delta E_{1/2}=4435\pm4\,\text{MeV}\,.
\eea
However in $B$ decays, both are kinematically forbidden, so they must be searched for in other decay channels.

Let us move to the analysis of $\tilde P_c(4337)$. We consider color-flavor is described by
\begin{equation}
    A^{udu}_{\alpha\beta\gamma}=u_{(\alpha}d_{\beta)}u_\gamma-u_{(\gamma}d_{\beta)}u_\alpha\
    \label{trentasei}
\end{equation} 
The combination $A^{uud}=0$ and $A^{udu}_{\alpha\beta\gamma}=-A^{duu}_{\alpha\beta\gamma}$. 
From this we have that the $ud$ pair is found in a symmetric color pairing, $J^{ud}_S$, whereas  $uu$ in in a  anti-symmetric  $J^{uu}_A$ pairing.  This allows to determine the shifts
\begin{equation}
    \begin{split}
        &\Delta E_{3/2}=2\,J_S^{ud}+J_A^{uu}=2\,J_S^{qq}+J_A^{qq}=17\pm 6~\text{MeV}\,,\\
        &\Delta E_{1/2}=72.7\pm 4 ~\text{MeV}\,.
    \end{split}
    \label{eq:massa_p_sbagliati}
\end{equation}
We will asume that  $\widetilde{M}_0=\widetilde{M}_0^s-\Delta_s=4259\pm6\,\text{MeV}$, with the same $\Delta_s$ computed above.\\
The mass spectrum is:
\begin{eqnarray}
        \tilde M_0+\Delta E_{3/2} &=& 4276\pm12~\text{MeV}=M_{\tilde P_c^\prime(3/2) }\,,\notag\\
        \tilde  M_0+\Delta E_{1/2} &=& 4332\pm7~\text{MeV} \longrightarrow \tilde P_c^+(4337)\,,
        \label{eq:Mass_Pcs2}\\
        \tilde  M_0-\Delta E_{1/2}&=&4187\pm7 ~\text{MeV}=M_{\tilde P_c^{\prime\prime}(1/2) }\,.\notag
\end{eqnarray}
We want to point out that, contrary to previous cases, for the spectrum of $\widetilde{P}_c$, we did not take the observed particle $\widetilde{P}_c(4337)$ as a reference point to obtain other predictions. Instead, we relied on the spectrum of $\widetilde{P}_{cs}$ and on the previously determined couplings.

\section{\texorpdfstring{$\tilde P$}{Ptilde} production channels}
Assuming that the color-flavor of the spectator $ud$ pair is mainteined in the final state, the process $\Lambda_b^0\to\Tilde{P}_c\,K^-$  is expected  to be suppressed, as suggested by observation, because in the diagram in Figure~\ref{fig:decay} the initial $[ud]$ diquark is in a color-flavor symmetric configuration as opposite to the $ud$  pair in $\tilde P$, which has to be in a antisymmetric configuration as concluded in~Eq.~\eqref{trentasei}. 

The assumption above suggests in turn that since baryons belonging to the sextet, namely $\Sigma_b,\,\Xi_b^\prime$ and $\Omega_b$, carry a color-flavor antisymmetric light quark pair, the process $\Xi_b^{\prime-}\to\Tilde{P}_{cs}\,K^-$ should be allowed. The $\Sigma_b$ baryon predominantly decays strongly, whereas observed pentaquarks are produced mainly in weak decays. We summarize in Table \ref{tab:new_decays} the allowed $\tilde P$ production processes initated by a baryon. We divide them into dominant and Cabibbo suppressed. These processes have large phase spaces allowing the production of the pentaquarks $\Tilde{P}_{cs}^\prime$ and $\Tilde{P}_{cs}^{\prime\prime}$.

\begin{table}[h]  
    \tabulinesep=1.mm
    \centering
    \begin{tabu}{|c||c|}
    \hline
        Dominant & Cabibbo suppressed \\
    \hline\hline
        - & $\Xi_b^{\prime0}\to\Tilde{P}_c\,K^-$\\
    \hline\hline
        {\setlength{\jot}{1.5 mm}
        $\!\begin{aligned}
           &\Xi_b^{\prime-}\to\Tilde{P}_{cs}\,K^-\\
           &\Xi_b^{\prime0}\to\Tilde{P}_{cs}\,\overline{K}^0\\
       \end{aligned}$} & 
       {\setlength{\jot}{1.5 mm}
       $\!\begin{aligned}
           &\Xi_b^{\prime-}\to\Tilde{P}_{cs}\,\pi^- \\
           &\Xi_b^{\prime0}\to\Tilde{P}_{cs}\,\pi^0
       \end{aligned}$}\\
       - & \,\,\,$\Omega_b^{-}\to\Tilde{P}_{cs}\,K^-$\\
    \hline
    \end{tabu}
    \caption{Possible new decay processes of sextet baryons into a pentaquark and a meson. We divide the processes into dominant and Cabibbo suppressed.}
    \label{tab:new_decays}
\end{table}

\begin{figure}[t]
        \subfloat[\label{fig:sdecay}]{\includegraphics[width=0.43\textwidth]{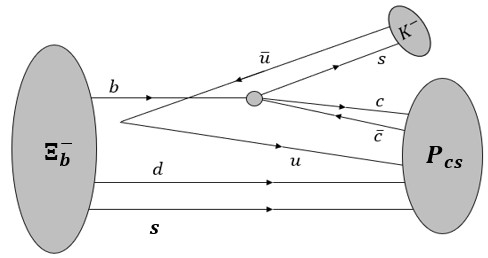}}
        \qquad
        \subfloat[\label{fig:newdecay}]{\includegraphics[width=0.43\textwidth]{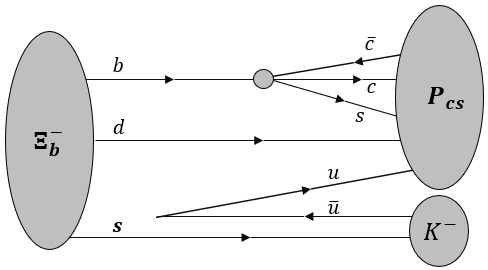}}\\
        \caption{Possible diagrams for the decay $\Xi_b^-\to P_{cs}\,K^-$. The diagram on the left is analogous to the one in Fig. \ref{fig:decay} and is suppressed for the $\Tilde{P}_{cs}$. The diagram on the right, however, does not have an analogue in the case of the pentaquarks $P_c$ but must be suppressed too since the process $\Xi_b^-\to \Tilde{P}_{cs}\,K^-$ is not observed.}
        \label{fig:new_decay}
\end{figure}

The decay $\Xi_b^-\to\Tilde{P}_{cs}(4338)K^-$ has not been observed either. If we consider the decay process of $\Xi_b^-$, we have another diagram, as shown in Fig. \ref{fig:newdecay}, which is not possible in the case of $\Lambda_b^0$. In absence of experimental evidence, we conclude that both diagrams in Fig.~\ref{fig:sdecay} and \ref{fig:newdecay} have to be suppressed. In \cite{SU3F} Han et al. have studied the production of pentaquarks through weak decays of $b$-baryons in the $\SU(3)_F$ symmetry. Following~~\cite{SU3F} it can be found that, if the process in Fig. \ref{fig:newdecay} is suppressed, then
\begin{equation}
    \mathcal{R}=\frac{|\mathcal{A}(\Lambda_b^0\to P_c(4312)\,K^-|^2}{|\mathcal{A}(\Xi_b^-\to P_{cs}(4338)\,K^-)|^2}=6.
\end{equation}
If this is not confirmed by data, a strong SU(3)$_F$ breaking effect would emerge, and the $\Xi$ should be effective at producing $\tilde{P}$ pentaquarks.

\section{Summary}
\label{sec:summary}
We summarize in Tables~\ref{tab:J_coupl} and~\ref{tab:Pentaquark_final} the values of the $J$ couplings and the masses of the observed and predicted Pentaquarks.
\begin{table}[hb!]
    \centering
    \begin{tabular}{|c|c|c|c|}
    \hline
        & Symmetric [MeV] & Antysimmetric [MeV] & No symmetry [MeV]\\
        \hline\hline
        $J^{qq}$ & $29.9^{+2.5}_{-2.8}$ & $-42.8^{+2.4}_{-1.6}$ & -  \\
        $J^{qs}$ & $17.9\pm 2$ & $-25.7\pm 2$ & $-3.9\pm 2$\\
        \hline
    \end{tabular}
    \caption{Couplings between light quarks. The ratio $S/A$ is the same for strange and non-strange pentaquarks $S/A\simeq -0.7\pm 0.1$. }
    \label{tab:J_coupl}
\end{table}
\begin{table}[ht!]
    \renewcommand{\arraystretch}{1.2}
    \centering
    \begin{tabular}{|c|c||c|c|}
    \hline
         & Mass [MeV] & & Mass [MeV] \\
    \hline \hline
      $P_c(4312)$ & $(4311.9^{+7}_{-0.9})$ & $P_{cs}(4459)$ & $ (4458.8^{+6}_{-3.1})$\\
      $P_c(4440)$ & $(4440.0^{+4}_{-5})$ & $\mathbf{P}_{cs}^\prime$ & $4541\pm 6$\\
      $P_c(4457)$ & $(4457.3^{+7}_{-1.8})$ & $\mathbf{P}_{cs}^{\prime\prime}$ & $4565\pm 6$ \\
      \hline\hline
      $\mathbf{\widetilde{P}_c^{\prime\prime}}$ & $4187\pm7$& $\widetilde{P}_{cs}(4338)$ & $(4338.2\pm0.8)$ \\
      $\mathbf{\widetilde{P}_c^\prime}$ & $4276\pm12$&$\mathbf{\widetilde{P}_{cs}^\prime}$ & $4387\pm4$\\
      $\widetilde{P}_c(4337)$ & $4332\pm 7~(4337^{+7\,\, +2}_{-4\,\, -2})$ & $\mathbf{\widetilde{P}_{cs}^{\prime\prime}}$ & $4435\pm4$\\
      \hline
    \end{tabular}
    \caption{In this table we summarize all the masses of the pentaquarks. Pentaquarks in boldface, $\bf P$, are predictions: six particles are predicted. Experimental values are in parentheses and are taken as input to obtain predictions on the $J$ couplings and masses of the pentaquarks $\mathbf{P}$. An exception is the $\widetilde{P}_c(4337)$ for which we have both the prediction and the experimental value (see Sec.~\ref{ptilde}). Each triplet is ordered from top to bottom with $J=1/2,3/2,1/2$.}
    \label{tab:Pentaquark_final}
\end{table}

Up to this point we have not discussed the spin of the $c\bar c$ pair. All the considerations made so far would hold for both $S_{c\bar c}=0,1$ although  we have been tacitly assuming $S_{c\bar c}=0$.   The $J/\psi-\eta_c$ mass splitting, $\delta=283$~MeV, would   suggest higher multiplets in the 300~MeV mass span. The splittings due to $\bm S_q\cdot \bm S_{c,\bar c}$ couplings are known to be indeed hyperfine, giving effects on the spectrum difficult to resolve~\cite{pill}. 

Pentaquarks made of diquarks were first considered in~\cite{Maiani:2015vwa} and in~\cite{Maiani:2015iaa},\cite{Ali:2019clg}. For a review on the use of diquarks in exotic spectroscopy see~\cite{Esposito:2016noz} and~\cite{Ali:2019roi}. Pentaquarks as antiquark-diquark-diquark systems have also  been considered recently in~\cite{Semenova:2019gzf}. Since the time of those papers, the experimental situation of the observed spectrum has changed qualitatively and the picture we have now is more complicated. 
It is the aim of this letter to attempt a step in a unified description of the states we have by now, including  those whose final assessment is still work-in-progress, as  for the $P_{cs}$ and its possible partners.  

Differently from~\cite{bopenta}, we  consider the exchange interactions among light quarks rather the color $c(\bar c)-q$ interactions. These exchange interactions are supposed to generate the observed mass splittings. No such splittings are occurring in ordinary baryons. The exchange interaction leads us to combine color-flavor quantum numbers instead of flavor-spin, as done in~\cite{bopenta}. In contrast to \cite{bopenta}, we can provide a precise assignment for the spin and flavor of the observed and predicted pentaquarks, as well as a hint on their masses.

Compared to~\cite{xc} we consider only the $\mathbf{8}-\mathbf{8}$ component for the color part and not the singlet. In~\cite{xc}, it is assumed that $P_c(4312)$ has $J^P=3/2^-$. In this work, we conclude that $J^P=1/2^-$ without any prior assumptions on the spin of the observed pentaquarks. For the spin interaction, we parametrize the chromomagnetic interaction differently from~\cite{xc}, using the exchange interaction. The parameters of this interaction are fitted from the observed spectrum rather than taken from the baryonic spectrum as done in the other work. In tab.~\ref{tab:comp}, we compare the spin assignment obtained in~\cite{xc} with that of this work. We see that there is a discrepancy in the spins of the $P_c$ pentaquarks for which we have exactly the opposite assignment.
\begin{table}[t]
    \centering
    \begin{tabular}{|c|c|c|c|c|c|c|}
        \hline
        & $P_c(4312)$ & $P_c(4440)$ & $P_c(4457)$ & $\Tilde{P}_c(4337)$ & $P_{cs}(4459)$ & $\Tilde{P}_{cs}(4338)$\\
        \hline\hline
        This work & $1/2$ & $3/2$ & $1/2$ & $1/2$ & $1/2$ & $1/2$ \\
         S. Y. Li et al. \cite{xc}& $3/2$ & $1/2$ & $3/2$ & $1/2$ & $1/2$ & $1/2$\\
         \hline
    \end{tabular}
    \caption{Comparison between the spin predicted in this work and in~\cite{xc} for the observed pentaquarks. The spin of the $P_c(4312)$ in the second row is an assumption and the other values are calculated by treating this particle as the reference state.}
    \label{tab:comp}
\end{table}
As for the work in \cite{mit}, they employ the MIT bag model and the Young-Yamanouchi basis for the wavefunction, so they combine the color and spin quantum numbers. The spin interaction is considered as in \cite{xc}, using chromomagnetic coefficients. Additionally, they find a lower bound for the masses of $c\Bar{c}uud$ pentaquarks, suggesting that the $P_c(4312)$ and $P_c(4337)$ cannot be compact states but rather molecular ones. This work, like \cite{xc}, does not seem to support this hypothesis.

Overall, the cited works do not address the issue of the suppression of pentaquarks which we have called $\tilde{P}$ in this article, in the decays of $\Lambda_b$ and $\Xi_b$ baryons, which is experimentally observed.

The Fermi statistics of the light color-octet cloud $qqq$ bound to the compact $c\bar c$ core  to form a color singlet, together with restrictions on the signs of $J$ couplings and the ratio $S/A$, repulsive/attractive coupling ratio in a color pair, allows to accomodate the observed $P_c(4312)$, $P_c(4440)$ and $ P_c(4457) $ and predict the full strange triplet $P_{cs}(4459), P_{cs}^\prime,P_{cs}^{\prime\prime}$,   
as long as $M_{P_{cs}}-M_{P_c}\simeq M_\Lambda-M_p$. Both triplets are preferred in the spin ordering  $J=1/2,3/2,1/2$ for increasing mass.  The states observed in association with anti-protons, $\tilde P_c$ and $\tilde P_{cs}$ may also be accompained by partners. We identified two states, with no strangeness, dubbed $\tilde P_c^\prime, \tilde P_c^{\prime\prime}$ having  $J=3/2,1/2$ respectively. The difference between the $P$ and the $\tilde P$ series is traced back to different color-flavor organization quarks, $S$ for the the $P$ and $A$ for the $\tilde P$.  

\section*{Acknowledgements}
We would like to thank Vanya Belayev and his LHCb collaborators for providing many useful information and comments on the manuscript.

\appendix
\section{Appendix: Three fermions exchange interaction}
\label{app:6jwigner}
In this appendix, we want to provide a demonstration of~\eqref{eq:shift_1/2} and~\eqref{eq:shift_3/2} using the 6j-Wigner symbols. The interaction we will study is given by~\eqref{exchange}:
\begin{equation}
    V=-\sum_{\text{pairs}}J_{ab}\left(\frac{1}{2}+2\mathbf{S}_a\cdot\mathbf{S}_b\right).
\end{equation}
$V$ commutes with $\mathbf{S}_{tot},\,\mathbf{S}_{tot}^z$, and $\mathbf{S}_i^2$ with $i=1,2,3$. Therefore, we can take the basis of the space as the kets $|S_{12},S_3;S_{tot},S_{tot}^z\rangle$, where $\mathbf{S}_{12}=\mathbf{S}_1+\mathbf{S}_2=0,1$. Explicitly,
\begin{equation}
    |1,\frac{1}{2};\frac{3}{2},+\frac{3}{2}\rangle,\,|1,\frac{1}{2};\frac{3}{2},+\frac{1}{2}\rangle,\,|1,\frac{1}{2};\frac{3}{2},-\frac{1}{2}\rangle,\,|1,\frac{1}{2};\frac{3}{2},-\frac{3}{2}\rangle\,;
\end{equation}
\begin{equation}
    |1,\frac{1}{2};\frac{1}{2},+\frac{1}{2}\rangle,\,|0,\frac{1}{2};\frac{1}{2},+\frac{1}{2}\rangle,\,|1,\frac{1}{2};\frac{1}{2},-\frac{1}{2}\rangle,\,|0,\frac{1}{2};\frac{1}{2},-\frac{1}{2}\rangle\,.
\end{equation}
The matrix associated with $V$ in this basis is a 2-block matrix: one for the total spin $3/2$ with size $4\times4$ and one for the total spin $1/2$ with the same size. Also, since there is no dependence on $\mathbf{S}_{tot}^z$, the eigenvalues are independent of this value. Thus, the first block is a multiple of the identity, while the second one can be decomposed into two $2\times 2$ blocks, one for $\mathbf{S}_{tot}^z=+\frac{1}{2}$ and one for $\mathbf{S}_{tot}^z=-\frac{1}{2}$ that have the same eigenvalues. We can, therefore, omit $\mathbf{S}_{tot}^z$ in the notation as it does not affect the eigenvalue calculation.
\subsection{\texorpdfstring{$\mathbf{S}_{tot}=3/2$}{S}}
For the total spin $3/2$, there is only one matrix element to calculate:
\begin{equation}
    \langle 1,\frac{1}{2};\frac{3}{2}|\,V\,|1,\frac{1}{2};\frac{3}{2}\rangle=-\frac{1}{2}\left(J_{12}+J_{13}+J_{23}\right)-\sum_{\text{pairs}}\langle 1,\frac{1}{2};\frac{3}{2}|\,2\mathbf{S}_a\cdot\mathbf{S}_b\,|1,\frac{1}{2};\frac{3}{2}\rangle\,.
    \label{eq:H_3/2}
\end{equation}
The matrix elements on the right-hand side are easily calculable. Starting with the case $a=1$ and $b=2$:
\begin{equation}
    2\mathbf{S}_1\cdot\mathbf{S}_2=\mathbf{S}_{12}^2-\frac{3}{2}=S_{12}\,(S_{12}+1)-\frac{3}{2}=\begin{cases}
        -\frac{3}{2} &\text{if } S_{12}=0\\
        \,\,\,\,\frac{1}{2} &\text{if } S_{12}=1
    \end{cases}\,.
\end{equation}
So
\begin{equation}
    \langle 1,\frac{1}{2};\frac{3}{2}|\,2\mathbf{S}_1\cdot\mathbf{S}_2\,|1,\frac{1}{2};\frac{3}{2}\rangle=\frac{1}{2}J_{12}\,.
    \label{eq:H_3/2_12}
\end{equation}
To calculate the other two matrix elements, it is necessary to change the basis and switch to $|S_{23},S_1;S_{tot}\rangle$ or $|S_{13},S_2;S_{tot}\rangle$. In the next section, we will see how to use 6j-Wigner symbols to make this change of basis. For spin $3/2$, there is no need because to obtain $\mathbf{S}_{tot}=\frac{3}{2}$, the only way is for the spin $\mathbf{S}_{ab}=1$ for any pair $a,b$. Therefore,
\begin{equation}
    \langle 1,\frac{1}{2};\frac{3}{2}|\,2\mathbf{S}_2\cdot\mathbf{S}_3\,|1,\frac{1}{2};\frac{3}{2}\rangle=\frac{1}{2}J_{23}\,,
    \label{eq:H_3/2_23}
\end{equation}
\begin{equation}
    \langle 1,\frac{1}{2};\frac{3}{2}|\,2\mathbf{S}_1\cdot\mathbf{S}_3\,|1,\frac{1}{2};\frac{3}{2}\rangle=\frac{1}{2}J_{13}\,.
    \label{eq:H_3/2_13}
\end{equation}
This result was predictable considering that spin-$\frac{3}{2}$ states are completely symmetric under the exchange of any pair of particles.\\
Combining \eqref{eq:H_3/2} with \eqref{eq:H_3/2_12} - \eqref{eq:H_3/2_23} and \eqref{eq:H_3/2_13},
\begin{equation}
    \Delta E_{3/2}\equiv\langle 1,\frac{1}{2};\frac{3}{2}|\,V\,|1,\frac{1}{2};\frac{3}{2}\rangle=-\frac{1}{2}\left(J_{12}+J_{13}+J_{23}\right)-\frac{1}{2}\left(J_{12}+J_{13}+J_{23}\right)=-J_{12}-J_{13}-J_{23}\,.
\end{equation}
\subsection{\texorpdfstring{$\mathbf{S}_{tot}=3/2$}{S}}
The eigenvalues of $\mathbf{S}_{tot}=\frac{1}{2}$ are obtained by diagonalizing the $2\times 2$ matrix:
\begin{equation}
    V=\begin{pmatrix}
        \langle 1|\,V\,|1\rangle & \langle 1|\,V\,|0\rangle \\
        \langle 0|\,V\,|1\rangle & \langle 0|\,V\,|0\rangle
    \end{pmatrix}\,.
\end{equation}
To calculate the matrix elements, we need to evaluate terms of the form:
\begin{equation}
    \langle S_{12},S_3;S_{tot}|\,2\mathbf{S}_{i}\cdot\mathbf{S}_{j}\,|S^\prime_{12},S_3;S_{tot}\rangle \quad\quad (i,j)\neq(1,2)\,.
    \label{eq:matrix_el}
\end{equation}
A transition from one coupling scheme to another is performed by a unitary transformation which relates the states with the same total spin $\mathbf{S}_{tot}$. From \cite{6j}, the unitary transformation we are looking for is
\begin{equation}
    \langle S_{12},S_3;S_{tot}|S_{23},S_1;S_{tot}\rangle\equiv(-1)^{S_1+S_2+S_3+S_{tot}}\sqrt{(2S_{12}+1)(2S_{23}+1)}\left\{
    \begin{matrix}
        S_1 &S_2 & S_{12} \\
        S_3 & S_{tot} & S_{23} \\
    \end{matrix}
\right\}\,,
\label{eq:6jS_23}
\end{equation}
where the last term is called 6j-Wigner symbol.\\
From this relation, follows:
\begin{equation}
    \langle S_{12},S_3;S_{tot}|S_{13},S_2;S_{tot}\rangle\equiv(-1)^{S_2+S_3+S_{12}+S_{13}}\sqrt{(2S_{12}+1)(2S_{13}+1)}\left\{
    \begin{matrix}
        S_2 &S_1 & S_{12} \\
        S_3 & S_{tot} & S_{13} \\
    \end{matrix}
\right\}\,,
\label{eq:6jS_13}
\end{equation}
\begin{equation}
    \langle S_{23},S_1;S_{tot}|S_{13},S_2;S_{tot}\rangle\equiv(-1)^{S_1+S_{tot}+S_{23}}\sqrt{(2S_{23}+1)(2S_{23}+1)}\left\{
    \begin{matrix}
        S_1 &S_3 & S_{13} \\
        S_2 & S_{tot} & S_{23} \\
    \end{matrix}
\right\}\,,
\end{equation}
which collectively provide all possible relations for transitioning from one basis to another. Specifically, the 6j-Wigner symbols are defined from Clebsh-Gordan coefficients as:
\begin{eqnarray}
    \sum_{\substack{m_{1},m_{2},m_{3},\\m_{12},m_{13}}} C^{jm}\left(j_{12}m_{12},j_3m_3\right)C^{j_{12}m_{12}}\left(j_{1}m_{1},j_2m_2\right)C^{jm}\left(j_{23}m_{23},j_1m_1\right)C^{j_{23}m_{23}}\left(j_{2}m_{2},j_3m_3\right)\nonumber\\=(-1)^{j_1+j_2+j_3+j}\sqrt{(2j_{12}+1)(2j_{23}+1)}\left\{
    \begin{matrix}
        j_1 &j_2 & j_{12} \\
        j_3 & j & j_{23} \\
    \end{matrix}
\right\}\,.
\end{eqnarray}
Now, let's calculate the matrix element \eqref{eq:matrix_el} explicitly. Starting with the term $2\mathbf{S}_2\cdot\mathbf{S}_3$, we use \eqref{eq:6jS_23} to transition to a more convenient basis:
\begin{multline}
    \langle S_{12},S_3;S_{tot}|\,2\mathbf{S}_{2}\cdot\mathbf{S}_{3}\,|S^\prime_{12},S_3;S_{tot}\rangle=\sum_{S_{23}}(-1)^{2(S_1+S_2+S_3+S_{tot})}\sqrt{(2S_{12}+1)(2S^\prime_{12}+1)}\\(2S_{23}+1)\left\{
    \begin{matrix}
        S_1 &S_2 & S_{12} \\
        S_3 & S_{tot} & S_{23} \\
    \end{matrix}
\right\}\left\{
    \begin{matrix}
        S_1 &S_2 & S^\prime_{12} \\
        S_3 & S_{tot} & S_{23} \\
    \end{matrix}
\right\}\left[S_{23}(S_{23}+1)-\frac{3}{2}\right]\,.
\end{multline}
For the term $2\mathbf{S}_1\cdot\mathbf{S}_3$, we use \eqref{eq:6jS_13}:
\begin{multline}
    \langle S_{12},S_3;S_{tot}|\,2\mathbf{S}_{1}\cdot\mathbf{S}_{3}\,|S^\prime_{12},S_3;S_{tot}\rangle= \sum_{S_{13}}(-1)^{2S_2+2S_3+2S_{13}+S_{12}+S^\prime_{12}}\sqrt{(2S_{12}+1)(2S^\prime_{12}+1)}\\(2S_{13}+1)\left\{
    \begin{matrix}
        S_2 &S_1 & S_{12} \\
        S_3 & S_{tot} & S_{23} \\
    \end{matrix}
   \right\}\left\{
    \begin{matrix}
        S_2 &S_1 & S^\prime_{12} \\
        S_3 & S_{tot} & S_{23} \\
    \end{matrix}
    \right\} \left[S_{13}(S_{13}+1)-\frac{3}{2}\right]\,.
\end{multline}
For a numerical example, let's calculate the matrix element $\langle 1|\,V\,|0\rangle$ for which contributions come only from the terms $2\mathbf{S}_2\cdot\mathbf{S}_3$ and $2\mathbf{S}_1\cdot\mathbf{S}_3$
\begin{multline}
    \langle1,\frac{1}{2};\frac{1}{2}|\,2\mathbf{S}_2\cdot\mathbf{S}_3\,|0,\frac{1}{2};\frac{1}{2}\rangle=\sqrt{3}\sum_{S_{23}=0,1}(2S_{23}+1)
    \left\{
    \begin{matrix}
        S_1 &S_2 & 1 \\
        S_3 & S_{tot} & S_{23} \\
    \end{matrix}
\right\}\left\{
    \begin{matrix}
        S_1 &S_2 & 0 \\
        S_3 & S_{tot} & S_{23} \\
    \end{matrix}
\right\}\left[S_{23}(S_{23}+1)-\frac{3}{2}\right]\\=
\sqrt{3}\left[\left(\frac{1}{2}\right)\left(-\frac{1}{2}\right)\left(-\frac{3}{2}\right)+3\left(\frac{1}{6}\right)\left(\frac{1}{2}\right)\left(\frac{1}{2}\right)\right]=\frac{\sqrt{3}}{2}\,;
\end{multline}
\begin{multline}
    \langle1,\frac{1}{2};\frac{1}{2}|\,2\mathbf{S}_1\cdot\mathbf{S}_3\,|0,\frac{1}{2};\frac{1}{2}\rangle=(-1)\sqrt{3}\sum_{S_{13}=0,1}(2S_{13}+1)
    \left\{
    \begin{matrix}
        S_2 &S_1 & 1 \\
        S_3 & S_{tot} & S_{13} \\
    \end{matrix}
\right\}\left\{
    \begin{matrix}
        S_2 &S_1 & 0 \\
        S_3 & S_{tot} & S_{13} \\
    \end{matrix}
\right\}\\\left[S_{13}(S_{13}+1)-\frac{3}{2}\right]=
-\sqrt{3}\left[\left(\frac{1}{2}\right)\left(-\frac{1}{2}\right)\left(-\frac{3}{2}\right)+3\left(\frac{1}{6}\right)\left(\frac{1}{2}\right)\left(\frac{1}{2}\right)\right]=-\frac{\sqrt{3}}{2}\,.
\end{multline}
\textbf{NOTE:} There is a minus sign between the two matrix elements due to the fact that in \eqref{eq:6jS_13} the exponent of the factor $-1$ depends on the spins $S_{12}$ and $S_{13}$.\\
The complete calculation leads to the matrix
\begin{equation}
    V=\begin{pmatrix}
        -J_{12}+\frac{1}{2}\left(J_{23}+J_{13}\right) & \frac{\sqrt{3}}{2}\left(J_{13}-J_{23}\right) \\
        \frac{\sqrt{3}}{2}\left(J_{13}-J_{23}\right) & J_{12}-\frac{1}{2}\left(J_{23}+J_{13}\right) 
    \end{pmatrix}\,,
\end{equation}
which has eigenvalues
\begin{eqnarray}
    \Delta E_{1/2}^+&=&\sqrt{J_{12}^2+J_{23}^2+J_{13}^2-J_{12}J_{23}-J_{12}J_{13}-J_{23}J_{13}}\,,\\
    \Delta E_{1/2}^-&=&-\sqrt{J_{12}^2+J_{23}^2+J_{13}^2-J_{12}J_{23}-J_{12}J_{13}-J_{23}J_{13}}\,.
\end{eqnarray}

\newpage
\bibliographystyle{unsrt}

\end{document}